# The Physics of Solar Energetic Particles

**Donald V. Reames** (https://orcid.org/0000-0001-9048-822X )
(retired) Wilmington, NC 28409
dvreames@gmail.com

**Abstract**

Solar energetic particles (SEPs) are produced in two fundamental ways: at magnetic reconnection sites in solar jets and at collisionless shock waves driven by fast coronal mass ejections (CMEs). "Impulsive" SEP events, on open field lines from jets, have signature abundance enhancements of $^{3}$He and of increasingly heavy elements, and their outward streaming electrons drive type-III radio bursts. Similar acceleration for particles trapped on closed loops energizes solar flares. In contrast, fast, wide, CME-driven shocks accelerate seed ions from the ambient corona that grow resonant Alfvén waves as they stream outward. These waves can scatter and trap lower-rigidity ions near the shock, limiting outflow, and flattening low-energy spectra upstream at the "streaming limit." Downstream, a spatially-uniform "reservoir" of SEPs is shed by the expanding shock between it and the Sun. These trapped invariant SEP spectra decrease in intensity adiabatically as the volume of the reservoir expands. Ions from the reservoir can seed further acceleration in multi-shock events and energetic proton precipitation can prolong solar $\gamma$-ray emission. Shocks are often additionally seeded by residual impulsive ions which dominate the SEP heavy-ion abundances with their signature enhancements. As samples of the corona, SEP abundances also probe differences with the photosphere that depend upon the first ionization potential (FIP) of the elements.

# 1 Introduction

Solar energetic particle (SEP) events are bursts of electrons and ions from the Sun, with energies from ~100 keV to several GeV, lasting from hours to days. All must be accelerated in and above the low-density solar corona, where Coulomb collisions cannot stop them, although some electrons and ions later plunge downward to produce X-rays and γ-rays, respectively. We have come to understand that in the largest, most-energetic "gradual" events the SEPs result from collisionless shock waves, driven by fast coronal mass ejections (CMEs), after they rise above the tops of coronal magnetic loops.

Smaller "impulsive" events, characterized by SEPs with greatly enhanced ratios of $^{3}He/^{4}He$ and of increasingly heavy elements have been associated with solar jets where oppositely-directed magnetic field reconnection and resonant wave absorption take place on open[1] magnetic fields, although similar processes also occur on closed loops to energize solar flares. Furthermore, gradual CME-driven shocks sometimes pass across active regions where residual, impulsive, pre-enhanced SEP seed ions have been accelerated from many jets, so these can dominate the final SEP heavy-element abundances.

Recent years have brought new measurements of element abundances and energy spectra of SEPs, as well as their distributions in space and time. These have helped us understand not only the element abundances and temperatures in the solar corona, and the physical processes of SEP acceleration and transport, but also the extent of those processes in CME-driven shocks, jets, flares, and even magnetic switchbacks in the solar wind.

SEPs reveal a broad range of astrophysical processes involving physics we cannot easily study remotely. Solar X-rays and radio emission produced by electrons and provide background, while γ-ray lines identify ions in a few large flares, but the direct samples of SEP ions and their abundances extend this study to physical processes which are otherwise inaccessible. In this article we summarize what has been learned about how the physics of SEPs fits together, and the way that knowledge has expanded in recent years.

# 2 History

Solar energetic particles (SEPs) were first reported at their highest energies where GeV protons form a nuclear cascade through the atmosphere to produce a shower of particles seen as a ground-level enhancements (GLEs). These are seen above the background of similar energetic events produced by the galactic cosmic rays (GCRs). Forbush (1946) observed a sharp increase in ionizing radiation prior to a "Forbush decrease" in GCRs that we now know as magnetic exclusion of GCRs by a coronal mass ejection (CME). Ironically, the SEPs are accelerated at the shock wave driven by the same fast CME that excludes the GCRs. However, Forbush associated the SEPs with solar flares, the most conspicuous solar signal. It took many years to even observe CMEs and then to associate SEPs with fast, wide CMEs producing shock waves (Kahler et al. 1984), sometimes even CMEs with no associated flare.

[1] Maxwell's equations require that all magnetic field lines are actually closed; however, we call lines "open" that have been drawn far out from the Sun by CMEs.

Meanwhile, there were also ground-based radio observations that identified two types of radio bursts driven by energetic SEP electrons (Wild, Smerd, and Weiss 1963). The emission frequency of waves is the local plasma frequency, which varies as the square root of the plasma density, declining as the source moves away from the Sun. There are fast, type-III radio bursts, produced by impulsive non-relativistic (10 – 100 keV) electrons streaming out from sources near the Sun - always scatter free (Tan et al. 2011), and there are slower type-II bursts where the frequency drift rate indicates a source moving out at the speed of an interplanetary shock wave. Thus, Wild, Smerd, and Weiss (1963) suggested that there are impulsive electron sources near the Sun, and shock sources, which were expected to accelerate protons and other ions efficiently. Later, when ~40 keV electrons were observed at 1 AU in conjunction with ~40 keV X-ray emission and type-III bursts, Lin (1970) observed that electrons must sometimes also be trapped at the Sun to produce flares and sometimes be emitted on open field lines to produce type-III radio bursts. He pointed out that "the magnitude of the magnetic field necessary to confine these electrons is very small" (Lin 1970), an observation relevant to the distinction much later between flares and solar jets, which are the open-field SEP sources [e.g. Bučík 2020].

However, the suggestion of shocks, from 1963 and from early space observations, that these shocks and type II bursts were associated with SEPs, was buried for several decades in what was later called the "solar flare myth" [Gosling 1993,1994], i.e. that all the SEPs we see in space are accelerated in solar flares. Some early models involved SEP diffusion far around the corona from flares (across field lines) followed by subsequent slow release [e.g. Reid 1964]. Then there was the "birdcage model" [Newkirk and Wenzel 1978] where arcades of coronal loops, like the wires of a birdcage, spread SEPs from a flare across the Sun in some rigidity-independent way. These models and the growth of evidence against them are discussed by Reames [2021a]; among them, the broad longitude expanse of SEP events, and the ionization states of Fe, $Q_{Fe}$ = 11 – 15 at energies up to 600 MeV amu$^{-1}$ [Luhn et al. 1984, 1987; Leske et al. 1985; Tylka et al. 1985], are incompatible with the high temperatures in flares. Clear evidence for SEP acceleration by CME-driven shocks solidified with the 96% correlation found by Kahler et al. [1984] and it continues in modern multi-spacecraft studies that have associated the broad spatial distribution of SEPs with comparably broad shock waves observed over extremely wide longitude spans [Rouillard et al. 2011, 2012; Gopalswamy et al. 2012; Lario et al. 2006, 2013; Desai and Giacalone 2016; Cohen et al. 2017; Kouloumvakos et al. 2019; Reames 2023a, b, 2025]. The strong connection of SEPs with type-II radio bursts continues in non-GLEs with the emission from shocks that persist beyond ~3 solar radii at 14 MHz [Cliver, Kahler, and Reames 2004].

# 3 $^3$He-rich Events and Flares

Measurement of the particle composition of SEPs became a significant factor in reestablishing a two-source SEP model. Need for a new type of physics became evident when $^3$He-rich events were found. Most workers had been familiar with the ~10% $^3$He in GCRs, produced as a fragment in nuclear reactions of GCR $^4$He ions with interstellar H during their ~$10^7$-year lifetime. However, SEP events had $^3$He/$^4$He as large as 1.5 ± 0.1 [Serlemitsos and Balasubrahmanyan, 1975; Mason, 2007] vs. a ratio of ≈ $5 \times 10^{-4}$ in the solar wind, and there was no measurable deuterium, ruling out fragmentation. Gamma-ray lines showed nuclear reactions in flares that would produce secondary species [Ramaty and Murphy 1987; Murphy et al 1991; Kozlovsky, Ramaty, and Murphy 2002], but subsequent search for the secondary elements Li, Be, and B found them to be <$2 \times 10^{-4}$ relative to O [Teegarden et al. 1973; McGuire, von Rosenvinge and McDonald 1979; Cook, Stone, and Vogt 1984]; again, there were no nuclear fragments. Thus, SEPs did not come out of flares. Also, the unique enhancement of rare $^3$He was not produced by fragmentation, but by a resonant wave-particle mechanism of which many were suggested [Ibragimov and Kocharov 1977; Kocharov and Kocharov 1978, 1984; Fisk 1978; Varvoglis and Papadopoulis 1983; Weatherall 1984; Winglee 1989; Riyopoulos 1991; Temerin and Roth 1992; Fitzmaurice, Drake, and Swisdak 2024].

It was a surprise when these $^3$He-rich events were found to be associated [Reames, von Rosenvinge, and Lin 1985; Reames and Stone 1986] with the streaming electron events that drove the type-III radio bursts, once called "pure electrons" by Lin [1970]. This connected these bizarre $^3$He-rich events with the extremely common type-III radio bursts. These were not pure-electron events. In a model by Temerin and Roth [1992] these streaming electrons generate electromagnetic ion cyclotron (EMIC) waves that resonate with the gyrofrequency of $^3$He, mirroring in the magnetic field, to produce the enhancements, in analogy with the "ion conics" in the Earth's magnetosphere.

Gamma-ray spectral lines tell us that the accelerated "beam" in flares is also $^3$He rich [Mandzhavidze et al. 1999; Murphy et al. 2016]. Gamma-rays emitted from excited states of ions of the beam are Doppler shifted to produce broad spectral lines, in contrast with the narrow lines from the ambient plasma. The especially strong reaction $^{16}O$ ($^{3}He$, p) $^{19}F^{*}$ leads to the three γ-ray lines at 0.937, 1.04, and 1.08 MeV from the deexcitation of $^{19}F^{*}$, allowing the measurement of accelerated $^{3}He$ in flares. In the study of 20 large flares, some had $^{3}He/^{4}He \sim 1$, while all had $^{3}He/^{4}He > 0.1$. Like the trapped and free electrons, noted by Lin [1970], the energy of magnetic reconnection can accelerate particles on closed or open field lines, leading to flares or to solar jets where the SEPs can run away freely.

Significant differences in the energy spectra of $^{3}He$ and $^{4}He$ are produced by the depletion of the rare $^{3}He$ in the source [Liu et al. 2004, 2006; Mason et al. 2024]. This depletion results in a limit on the fluence of $^{3}He$ ions [Ho, Mason, and Roelof 2005] in SEP events.

Some ion spectra in these events roll over below 100 keV amu$^{-1}$ [Bučik et al. 2016] suggesting acceleration deeper in the corona. Association of long durations of power-law spectra with spreading EIT waves below 1 MeV amu$^{-1}$ [Bučik et al. 2016] may suggest shock-like acceleration in the $\delta V$ of the wave.

Impulsive SEPs often form a narrow distribution in space [Reames 1999]. Of 15 events studied recently by Ho et al. [2024], 7 could not be seen by another spacecraft <20° away, although 5 events were seen by three spacecraft within 40°. However, some of these had associated CMEs where shocks could spread the distribution.

A substantial amount of work has now established the sources of $^{3}He$-rich events to be mainly solar jets, often with associated CMEs [Kahler, Reames, and Sheely 2001; Nitta et al. 2006; Wang, Pick, and Mason 2006; Reames, Cliver, and Kahler 2014a; Bučík et al. 2018a, b, 2021; Bučik 2020; Zhang et al. 2023], although early technology had found no clearly associated CMEs [Kahler et al. 1985]. While magnetic reconnection in a jet liberates SEPs on open field lines, it also traps some on newly closed loops to produce flaring, i.e. jets always have nearby flares. Blowout jets [e.g. Archontis and Hood 2013] may be associated with larger impulsive SEP events

# 4 Element Abundances

## 4.1 Impulsive and Gradual Events

Early measurements of heavier elements had shown that Fe was also enhanced in $^{3}He$-rich events [Mogro-Campero and Simpson 1972] and subsequent measurements showed that enhancements increased with atomic number $Z$ [e.g. Mason et al. 1986] in impulsive SEP events, and it was soon found that daily abundance averages showed a bimodal pattern [Reames 1988; Reames and Ng 2004; Cliver and Ling 2007; Cliver 2016]; days with impulsive and gradual events were fairly well distinguished. Breneman and Stone [1985;

Tylka et al. 2000] found a power-law dependence of abundance enhancement or suppression upon the atomic-mass-to-charge ratio *A/Q* for gradual SEP events, eventually averaging to the coronal reference abundances discussed in Section 6 below. Reames, Meyer, and von Rosenvinge [1994] suggested a similar power-law dependence upon *A/Q* for the enhancements in impulsive SEP events, with *Q* values at ~3 MK, where He, C, and O were fully stripped to have *A/Q* ≈ 2 and Ne, Mg, and Si were in stable configurations with 2 orbital electrons, so Ne/O, Mg/O and Si/O could be similarly enhanced. If Ne, Mg, and Si were also fully ionized, as at flare temperatures of ~ 10 MK, they could not be relatively enhanced at all. Subsequently, measurements that extended the abundances to *Z* ~ 82 showed average enhancements by a factor of up to ~1000, both above [Reames 2000; Reames and Ng 2004; Reames, Cliver and Kahler 2014a] and below [Mason et al. 2004] ~ 1 MeV amu$^{-1}$. Figure 1a shows a histogram of the distribution that separates impulsive and gradual periods of abundance enhancements in Ne/O vs. Fe/O during 8-hr periods over 19 years, while Fig. 1b shows the average enhancement vs. *A/Q* for 111 impulsive events [Reames, Cliver, and Kahler 2014a].

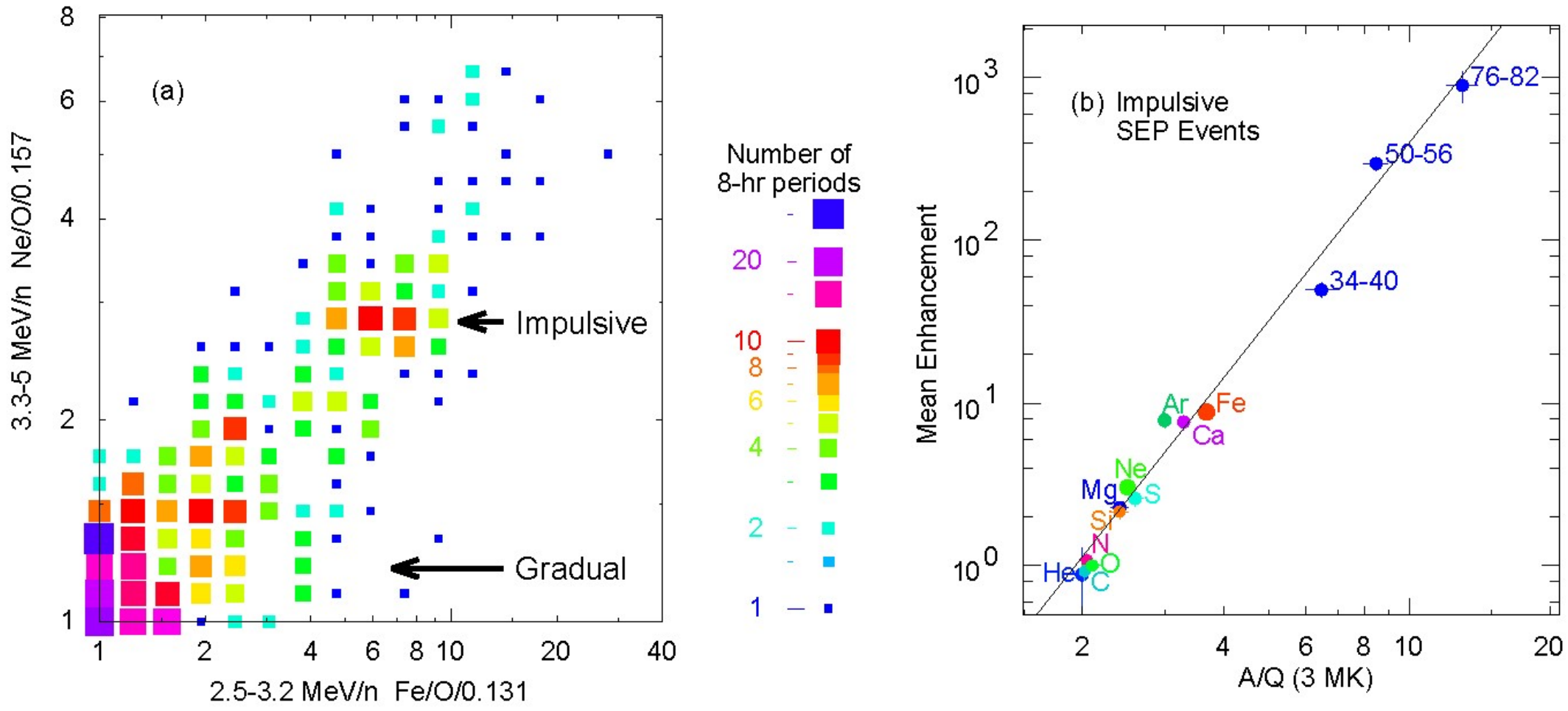

**Fig. 1** (**a**) A histogram shows the bimodal distribution of abundance enhancements of Ne/O vs. Fe/O measured in 8-hr periods over 19 years with impulsive and gradual regions indicated. (**b**) Shows a fitted power of the average enhancement vs. *A/Q* that is 3.64 ± 0.15 for 111 impulsive events at 3 – 10 MeV amu$^{-1}$.

The power-law abundance enhancements in impulsive SEP events are related primarily to the physics of acceleration in islands of magnetic reconnection (e.g. Drake et al. 2009). Power-law dependence in gradual events is related to reacceleration of this pre-enhanced impulsive seed population, in some events, or to rigidity-dependent scattering of ions during acceleration and transport. SEP scattering during transport can depend upon the initial state of the interplanetary plasma [e.g. Reames, Ng, and Berdichevsky 2001] or upon self-amplified waves [Ng and Reames 1994, 1995; Ng, Reames, and Tylka 2003, 2012].

## 4.2 Seed Populations

While it seemed simple to have only two populations of SEP events, impulsive and gradual, distinguished by their abundances, it soon became clear that residual ions from impulsive events could be reaccelerated by shock waves in gradual events when Mason, Mazur, and Dwyer [1999] found modest enhancements of $^{3}$He in a large gradual SEP event. Impulsive SEPs can produce a pre-enhanced seed population for shocks [Tylka et al. 2001, 2005; Desai et al. 2003; Tylka and Lee 2006; Sandroos and Vainio 2007] yielding the example shown below in Fig. 2h. Reames [2020] formalized this state with four types of SEP events.

- SEP1: "pure" impulsive SEPs from magnetic reconnection in jets (no fast shock).
- SEP2: Impulsive SEP1 ions reaccelerated by a local fast (>500 km $s^{-1}$) CME-driven shock from the same event.
- SEP3: A gradual event produced by a fast, wide CME-driven shock, where the high-*Z* SEPs are dominated by seed ions accumulated in its path in active regions with many small impulsive events.
- SEP4: A "pure" gradual event produced by a fast, wide CME-driven shock, dominated by accelerated seed ions from the ambient coronal plasma itself.

As expected from Fig. 1b, abundances in SEP1 events vary as a power of *A/Q*, although the power in individual events varies from 2 to 8, smaller events being steeper [Reames, Cliver, and Kahler 2014a]. Since the ion *Q* values vary with source electron temperature [e.g. Mazzotta et al. 1998; Post et al. 1977], one can choose the temperature that gives the best fit for power-law enhancement vs. *A/Q* for each impulsive SEP event [Reames, Cliver, and Kahler 2014b] or for gradual events [Reames 2016]. Surprisingly, the best-fit temperatures for the impulsive events fell in the narrow range 2.5 – 3.2 MK for 97% of the events. Subsequently, solar-jet source temperatures in $^{3}$He-rich events have been independently determined from EUV measurements to be ~2.5 MK [Bučík et al. 2021]. The charge states vs. temperature used for SEP studies were based upon Maxwellian electron distributions, but Lee et al. [2024] found that using kappa distributions produced similar temperatures and results. They noted that acceleration must occur early, before any significant heating.

At the other extreme are the classical gradual SEP4 events, with shock acceleration of ions from material of the ambient corona. With the exception of $^{3}$He, which may always be present at some low level [Mason, Mazur, and Dwyer 1999], we cannot always distinguish the presence of a tiny component of impulsive seed ions, but when the heavy ions in an event are dominated by an increasing power-law in *A/Q*, this enhancement likely comes from the seed population. In a SEP2 event, these seed ions may be reaccelerated by a local shock, driven by a sufficiently fast CME from the same jet that produced them [Kahler,

Reames, and Sheeley 2001]. SEP3 events are produced by a shock wave traversing an active region where the output from many impulsive jets has accumulated and may flow almost continuously as are commonly observed [Wiedenbeck et al. 2008; Bučík et al. 2014, 2015; Chen et al. 2015; Kouloumvakos et al. 2023; Lario et al. 2024;Reames 2024b]; often such regions produce sequences of SEP3 events [Fig. 3; Reames 2024b]. However, increasing abundances of heavy elements can also be produced in very large gradual SEP events when self-amplified Alfvén waves produced by particles streaming away from the shock (Section 5.2) cause scattering and trapping, from which high-rigidity (high *A/Q*) ions preferentially escape. This produces a correlation between observed energy spectra and abundances in SEP4 events [Reames 2022].

Three main abundance patterns in gradual SEP events are shown in Fig. 2 [Reames 2016, 2018b, 2024a]. The event on the left is a typical gradual event, with shock-accelerated seed ions of coronal abundances, where ions with higher rigidity (*A/Q*) scatter less and leak away more rapidly so enhancements usually decline with *A/Q*. The initial enhancement pattern in Fig. 2g is so flat that its best-fit temperature is poorly determined in Fig. 2d. The central event in Fig. 2 is a SEP3 event where already-enhanced impulsive seed ions dominate the heavy-element region above He, C and O (Fig. 2h). The event on the right in Fig. 2 is actually a GLE where the intense higher-energy protons have generated a streaming-limited plateau [Reames and Ng 2010; Ng, Reames, and Tylka 2003, 2012; Ng 2014] with intense scattering where the low-rigidity ions are trapped and the higher-rigidity ions penetrate out to the observer more easily (Fig. 2i) and the protons fit the same pattern; the abundances revert to the typical pattern, declining with *A/Q*, for the ions released behind the shock (see Section 5.2 below). The streaming limit has been rediscovered at low energies by Giacalone et al. [2023] in the "flat spectrum" upstream of a shock.

An occasional intermediate variant of the normal SEP4 pattern on the left in Fig. 2 occurs when the protons are trapped and suppressed but heavier ions are not; it is called "broken protons," and is described in Reames [2024b].

$^3$He/$^4$He is not actually a good measure for selecting impulsive events, since it is poorly defined, even varying by orders of magnitude with energy during a single event [e.g. Mason 2007]. Thus, Reames, Cliver, and Kahler [2014a] used the Fe/O abundance criterion shown in Fig. 1a to select "impulsive" events. However, in hindsight, it turns out that 40% of their 111 SEP events listed have associated CMEs with speeds > 500 km s$^{-1}$, so that many of them would be SEP2 or even SEP3 events by this standard. All are dominated by impulsive seed ions at high *A/Q*, but events with shocks would be expected to have broader spatial distributions (so they are more easily seen).

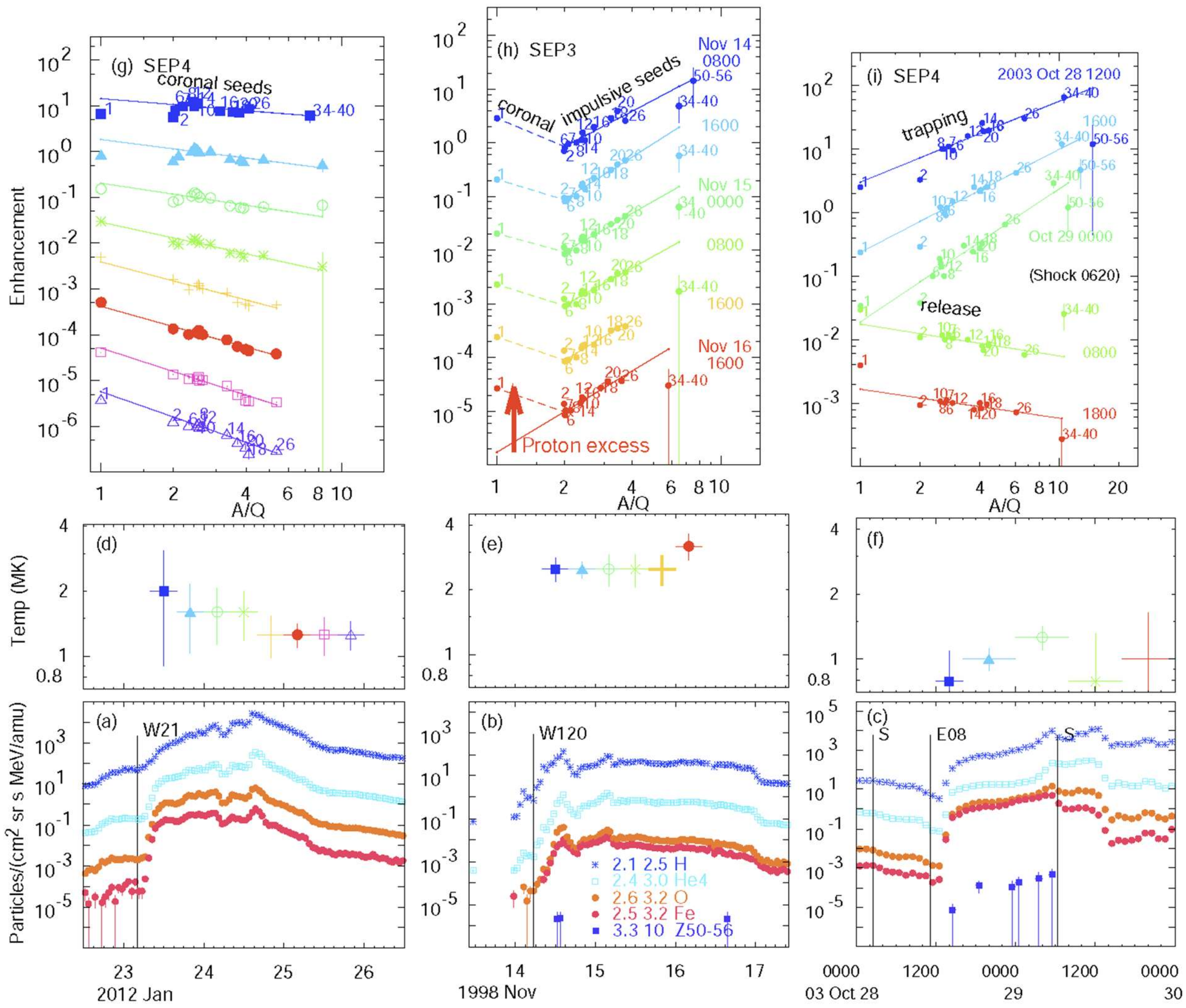

**Fig. 2** (**a**)-(**c**) show intensities of energy intervals (MeV amu$^{-1}$) of the ions listed for three gradual SEP events, (**d**)-(**f**) show the corresponding best-fit source temperatures and time intervals studied and (**g**)-(**i**) show power-law fits of element enhancements (labeled by Z) vs. A/Q for those time intervals (shifted by 10). The event on the left shows typical coronal seed intensities in (**g**). The central event is a SEP3 event with impulsive seed ions in (**h**). The right event (**i**) is a GLE where self-amplified waves preferentially trap lower-rigidity ions near the shock – especially protons – freeing all ions them behind it. Shock speeds are (**a**) 2175, (**b**) NA, and (**c**) 2459 km s$^{-1}$.

Plotting the He spectral index early in the events vs. the power of *A/Q*, Reames [2022] shows the distribution of SEPn of events in that space. SEP2 and SEP3 events are hard the distinguish, but large events with hard spectra are unlikely to occur from a single local jet, so these are probably SEP3. SEP4 events stretch from large events where hard spectra ensure intense scattering so enhanced heavy ions reach Earth, to smaller events with reduced scattering so heavy ion leak away early causing negative powers of *A/Q*. This behavior of SEP4, a joint result of acceleration and transport, and is evidence of self-generated waves.

SEP3 events seem to be much more prevalent during the active solar-cycle 23 (1996 – 2009). Of the 16 GLEs in that cycle, 6 were SEP3 events and 10 were SEP4 events (Reames 2024b). Figure 3 shows three successive events from a single active region as it rotates across the Sun, the last two events are SEP3 GLEs which have swept up the impulsive seed population. During the weaker cycle 24, Cohen, Mason, and Mewaldt [2017] studied element abundances in extensive multi-spacecraft events; only one (2011 August 4) of their 41 events was a SEP3 event. Probably both strong shocks and impulsive seed events are more common during stronger solar cycles. In a very active cycle, could all of the gradual events be SEP3?

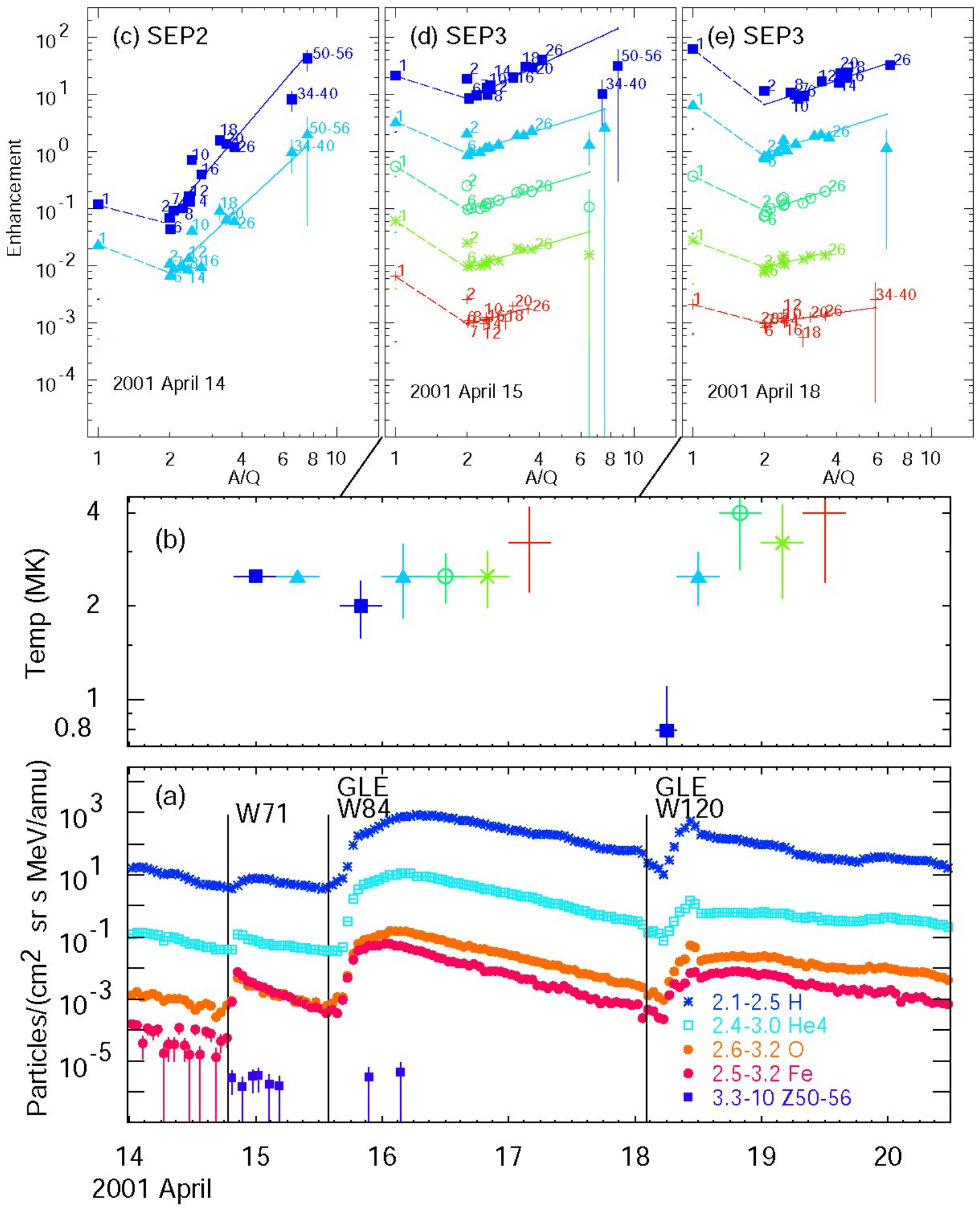

**Fig. 3** (**a**) shows intensities of the listed ions and MeV amu$^{-1}$ intervals during an SEP2 event and two gradual SEP3 events, both GLEs. Panel (**b**) shows the fitted temperatures ($\geq$ 2 MK) and time-interval durations. In (**c**), (**d**) and (**e**), abundance enhancements for ions, relative to O, are plotted with colors and symbols of the intervals in (**b**); indicative values of $Z$ are listed, and all $Z \geq 6$ ions are fitted vs. $A/Q$ (each interval shown shifted by a decade) with dashed lines drawn to the excess protons. CME speeds for the events at W71, W84, and W120 are 830, 1199, and 2465 km s$^{-1}$, respectively [Reames 2024b].

If SEP abundances depend upon the seed population, it should be possible for them to vary as a function of longitude along the shock depending upon the presence or lack of impulsive, suprathermal seed ions. Recently Xu et al. [2024] have found an event that, for the whole energy range of coverage, 0.1 – 10 MeV amu$^{-1}$, is Fe-rich at one spacecraft and Fe-poor at another, ~60° in longitude away.

## 4.3 Rarer Patterns of Impulsive Suprathermal Ions

On average, ions below 1 MeV amu$^{-1}$ have a similar power-law behavior to those above 1 MeV amu$^{-1}$; Mason et al. [2004] found enhancements varied as $(A/Q)^{3.26}$ on average, using $Q$ values at 3.2 MK, for ions with $6 \le A \le 200$, although individual events do show variations.

As spacecraft have made measurements closer to the Sun, it has been easier to study abundances in smaller and smaller impulsive SEP events that can only be seen below 1 MeV amu$^{-1}$. Recently, Bučík et al. [2025] have measured a small event with $^3$He/$^4$He = 75 ± 34 where the heavy ion enhancements increased unevenly to a peak at S, then declined to Fe; no ions were seen above 0.5 MeV amu$^{-1}$. They found the SEP source for this event to be a “straight tiny jet” on the edge of a coronal hole. As is typical of coronal holes, this source had a temperature of 1.5 – 1.7 MK. At this temperature, S can have $Q$ = 10 – 11 with $A/Q \approx 3$, allowing a second-harmonic resonance with the same waves that resonate with $^3$He, which has $A/Q$ = 1.5 [Mason et al 2024]. Previously, Mason et al. [2016] reported 16 SEP events in 16 years that had similar abundance peaks at Si and S. Abundances also vary with energy in these events. Apparently, in these small SEP events near cool coronal holes, second-harmonic wave resonances [e.g. Roth and Temerin 1997] dominate over power-law-producing reconnection [Drake et al. 2009] in determining the abundance pattern. These small localized events certainly lack fast CMEs that could produce SEP2 events and residual ions are too few to contribute seeds for subsequent higher-energy behavior. Other unusual enhancements, such as greatly increased abundances of ions with $A$ > 100 [Mason et al. 2023], can also occur at suprathermal energies in these small events. Perhaps we need a new category, SEP0, for tiny resonance-dominated events with varying source temperatures that are confined to energies below 1 MeV amu$^{-1}$.

In addition to flares and jets, there are other regions where magnetic reconnection occurs near the Sun. For example, magnetic switchbacks in the solar wind are sudden reversals in the field direction that return to the original direction in minutes. Switchbacks have been modeled [Zank et al. 2020] as an interchange reconnection between a coronal loop and open fields, similar to models for a simple jet [Shimojo and Shibata 2000], that propagates outward as a kink or compressed *S* shape. McDougall and Poduval [2025] have recently measured excess suprathermal “alphas” associated with switchbacks. Historically, nuclear alpha-particle radiation is purely $^4$He, never $^3$He, but the term has recently become used for events that are clearly $^3$He-rich [see Fig. 5 in Alnussirat et al. 2025]. Thus, switchbacks are a common consequence of reconnection, just like $^3$He-rich SEP events and type-III radio bursts? Actually, the SEPs would leave these reconnection events early but a few could be temporarily trapped in the switchback near the Sun.

# 5 Shock Acceleration

In diffusive shock acceleration (DSA), particles receive an increment of velocity as each scatters back-and-forth across the shock [e.g. Jones and Ellison 1991]. As they stream away from the shock at each new energy, they generate or amplify resonant Alfvén waves [Stix 1992; Melrose 1980] that scatter subsequent particles, trapping them so they gain additional energy and repeat the process. Particles, with velocity greater than the Alfvén speed $v_A$, resonate with waves of wave number $k$ with

$$k \approx B/\mu P, \qquad (1)$$

in the rest frame of the waves, where $P$ is the magnetic rigidity, i.e. momentum per unit charge of the particle, and $\mu$ is its pitch angle with respect to the magnetic field of vector $\boldsymbol{B}$. Bell [1978a, b] derived the equilibrium solution for this acceleration and Lee [1983] applied it to interplanetary shock acceleration; assuming $\mu \approx 1$ greatly simplified the solution so that each rigidity had its own resonant waves, with no cross coupling.

Shock acceleration of SEPs has now been demonstrated in many ways [Reames 1995b, 1999, 2013, 2021a; Reames, Kahler, and Ng 1997; Zank, Rice, and Wu 2000; Kahler 2001; Cliver, Kahler, and Reames 2004; Lee 2005; Cliver and Ling 2007, 2009; Rouillard et al. 2011, 2012; Gopalswamy et al. 2012, 2013a, b; Lee, Mewaldt, and Giacalone 2012; Desai and Giacalone 2016; Kouloumvakos et al. 2019; Li and Lee 2019]. A recent review of machine learning attempts to predict SEPs has been published by Kasapis et al. [2026]

## 5.1 Energetic Storm Particles (ESPs)

However, equilibrium solutions, which generate infinite power-law energy spectra in infinite time, are incomplete since real shocks move out into lower densities and lower magnetic fields where waves of given $k$ now resonate with particles of lower rigidity (Equation 1), so high-rigidity particles begin to leak away and the power-law spectrum rolls down. Upstream of the shock, particle scatter and intensities decrease with distance until there are too few resonant waves to contain those particles; this creates a region around the shock, bounded in energy and space, that was historically called an energetic storm-particle (ESP) event when it passed near Earth. The highest-energy particles trapped in the ESP region near the Sun are the first to leak away when the shock moves to lower $B$, but intense ESP events can still carry trapped protons of >100 MeV past 1 AU as shown for four large SEP events in Fig. 4. The highest-energy (~GeV) protons seen initially in these events have begun to leak away from the ESP soon after it was first formed near the Sun. ESP regions are best seen from events near central meridian where the most-persistent central nose of the shock survives to pass over the observer.

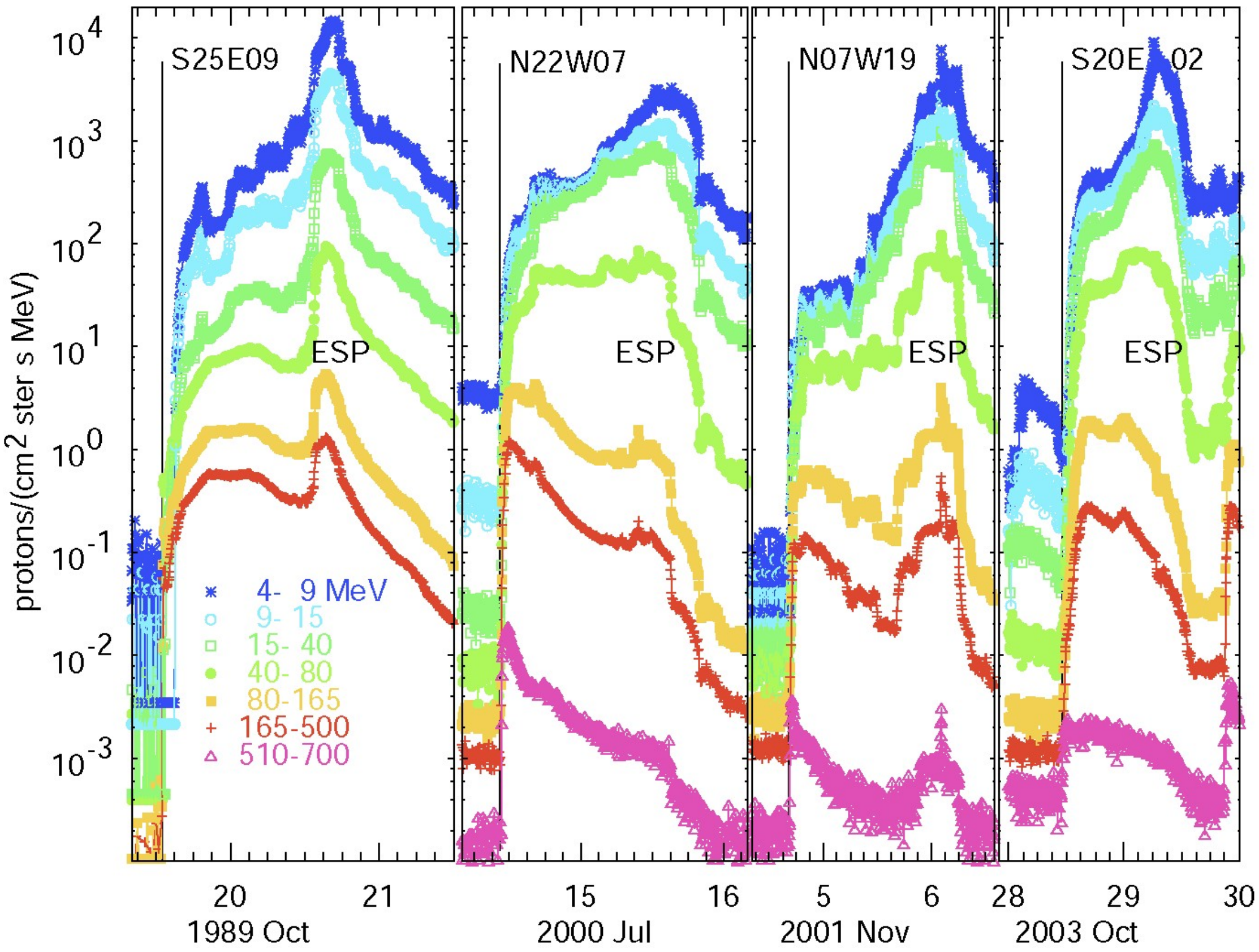

**Fig. 4** shows intensities at the NOAA/GOES spacecraft of protons in the listed energy intervals in four large SEP events where the ESP event peaks stand out around the shocks.

The radial evolution of an ESP event is suggested by the multi-spacecraft observations by *Helios*, IMP-8, and *Voyager* of 6 – 11 MeV protons shown in Fig. 5. At this relatively low energy, the intensity of this ESP event is rather well maintained during the expansion out to ≈ 2 AU but its width decreases; however, it is difficult to draw quantitative conclusions because of the possible variation with longitude. The ESP event shown here at *Helios 2* and IMP 8 actually extends behind the shock in the compression region ahead of the driving magnetic cloud (MC) of the CME [Burlaga et al. 1981]. Higher-energy SEP profiles for this event are shown in Reames [2023a, b]. There are some cases like this when multiple spacecraft are fortuitously positioned, but they rarely repeat; Reames [2023b] suggests a spacecraft configuration that could measure the same spatial distribution along CMEs, shocks, and SEPS for many events.

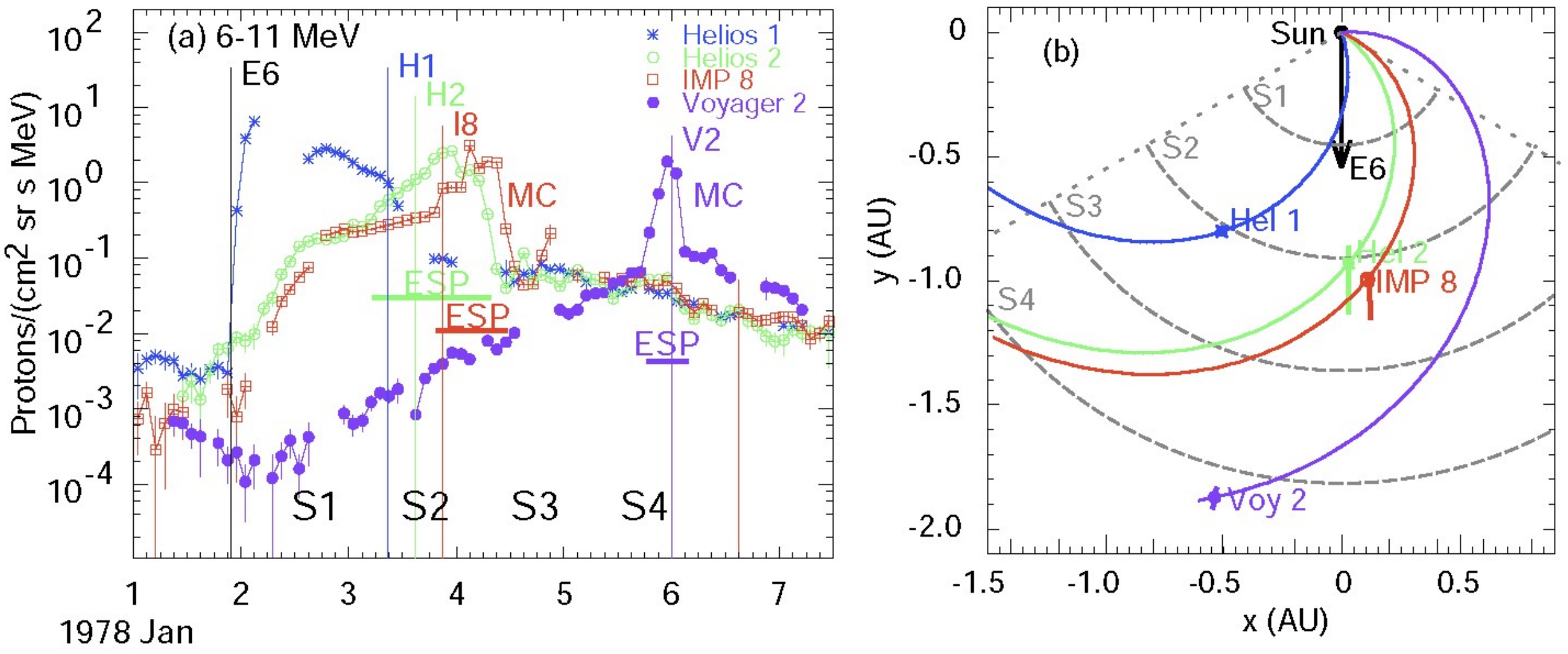

**Fig. 5** (**a**) observations of 6 – 11 MeV proton intensities vs. time are shown with shock-passage times and ESP regions noted for the listed spacecraft whose locations are shown in (**b**). MC denotes passage of the magnetic cloud of the CME, and approximate spatial widths of the ESP regions in (**b**) are mapped from corresponding durations in (**a**). Shock positions S1 – S4 in (**b**) are mapped approximately into times in (**a**).

Lee [2005] extended DSA theory, including ESP-event structure, analytically, by making simplifying assumptions. Vainio et al. [2014] undertook an extensive study of ESP event structure and developed a semi-empirical model that could be used as a realistic particle source. Kouloumvakos et al. [2025] used this ESP source to model the SEP spatial distribution and onset timing and found that perpendicular diffusion of the particles was no longer required. Perhaps this finally will end the misuse of perpendicular diffusion, once applied as a large adjustable parameter for shock-substitution models in the flare-myth era. However, there is a valid interest in lateral spreading of SEPs caused by random walk of field lines prior to an event [Jokipii and Parker, 1969; Li and Bian, 2023] that is best studied with shockless SEP1 events.

Study of the acceleration of high-energy SEPs considering a shock traversing a series of concentric spherical shells [Zank, Rice, and Wu 2000] and developed the PATH models [Zank, Li, and Verkhoglyadova 2007; Hu et al. 2018]. Rouillard et al. [2016], Afanasiev et al. [2018], Li and Lee [2019], and Kouloumvakos et al. [2020] showed that shock simulations easily produced GeV protons in GLEs, and Zhou et al. [2026] studied a 3D shock model.

After time-dependent wave generation was added to focused-transport theory [Ng and Reames 1994,1995; Ng, Reames, and Tylka 1999, 2003, 2012; Battarbee, Laitinen, and Vainio, 2011, Afanasiev et al. 2015, 2023], Ng and Reames [2008] were able to follow the evolution of fully time-dependent proton acceleration plus wave growth. This latter model found that a 2500 km $s^{-1}$ shock could accelerate protons to ~300 MeV in ~10 min. Including $\mu$ in the calculations showed that each new energy attained during acceleration began as a pancake distribution since ions at higher $P_{\mathrm{new}}$ with $P_{new}\,\mu$ at low $\mu$ could be initially scattered and trapped by waves previously generated by a lower $P_{old}$, i.e. when $\mu \leq P_{old}/P_{new} < 1$. Realistically, growth and subsequent absorption of the wave spectrum make the SEP scattering highly dependent upon both space and time [Ng, Reames, and Tylka 2003, 2012]. However, these fully time-dependent calculations require extensive computer time.

Non-relativistic electrons cannot resonate with Alfvén waves, so they tend to be accelerated on the flanks of the shock in the $\boldsymbol{V_{shock}} \times \boldsymbol{B}$ electric fields. Morosan et al. [2025] studied multi-frequency observations of type-II radio emission using the Low Frequency Array (LOFAR) and hard X-ray emission to map electron acceleration regions, often on the flanks of shocks. The spatial distribution of type-II emission may be quite complex. Lateral expansion of shock flanks is an important factor in the extent of the spatial distribution of SEPs in the heliosphere [Rouillard et al. 2012; Wijsen et al. 2025; Gopalswamy 2020].

In an analysis of 72 GLEs from 1942-2017, Cliver, Mekhaldi, and Muscheler [2020] determined that GLE spectra show a broad-scatter trend, from hard spectra for events originating behind the west limb of the Sun to soft spectra for GLEs associated with flares observed near solar central meridian. They attributed this trend to the longitude-dependent apportionment of the high-energy protons arriving at Earth between those accelerated by quasi-perpendicular CME-driven shocks propagating parallel to the solar surface and those accelerated at bow shocks driven radially outward by the CME, with quasi-perpendicular shocks producing a harder spectrum. This shock-related effect significantly broadens the longitude span of GLEs.

## 5.2 Upstream: Velocity Dispersion and the Streaming Limit

Far upstream of the shock, the first particles released, that have scattered minimally, are observed to arrive in inverse, order of their velocity $v$ in a time $t = L/v$ where $L$ is the pathlength from the shock along the magnetic field line. Figure 6a shows the ordered onset times for He of different energies in a large GLE event. Figure 6b shows onset times plotted as a function of $v^{-1}$ for the same event, where slope is the pathlength $L$ and the intercept is the earliest solar particle release (SPR) time from the shock at the Sun [Reames 2009a, b]. Figure 6c shows the location of the shock at SPR time vs. the magnetic connection longitude for several GLEs [Reames 2009b]. Well-connected GLE events often begin at ~2 solar radii, well after the onset of type-II radio emission, which occurs at ~1.5 solar radii on the flanks [Gopalswamy 2020], and after an associated flare is over [Tylka et al. 2003; Reames 2009a]. For smaller, non-GLE events, Cliver, Kahler, and Reames [2004] found type-II radio emission is seen at ~14 MHz, which corresponds to ~3 solar radii. Presumably, SPR time occurs as the shock emerges above closed coronal loops and exceeds the local Alfvén speed which is decreasing radially. Tan et al. [2013] compared path lengths of electrons and ions in GLEs and Kolympiris et al. [2023] compared them in other non-GLE

SEP events. The fitted parabola in Fig. 6c may be leading; onset on the flanks occurs when the shock *nose* is at ~ 6 $R_S$, not the magnetic connection point.

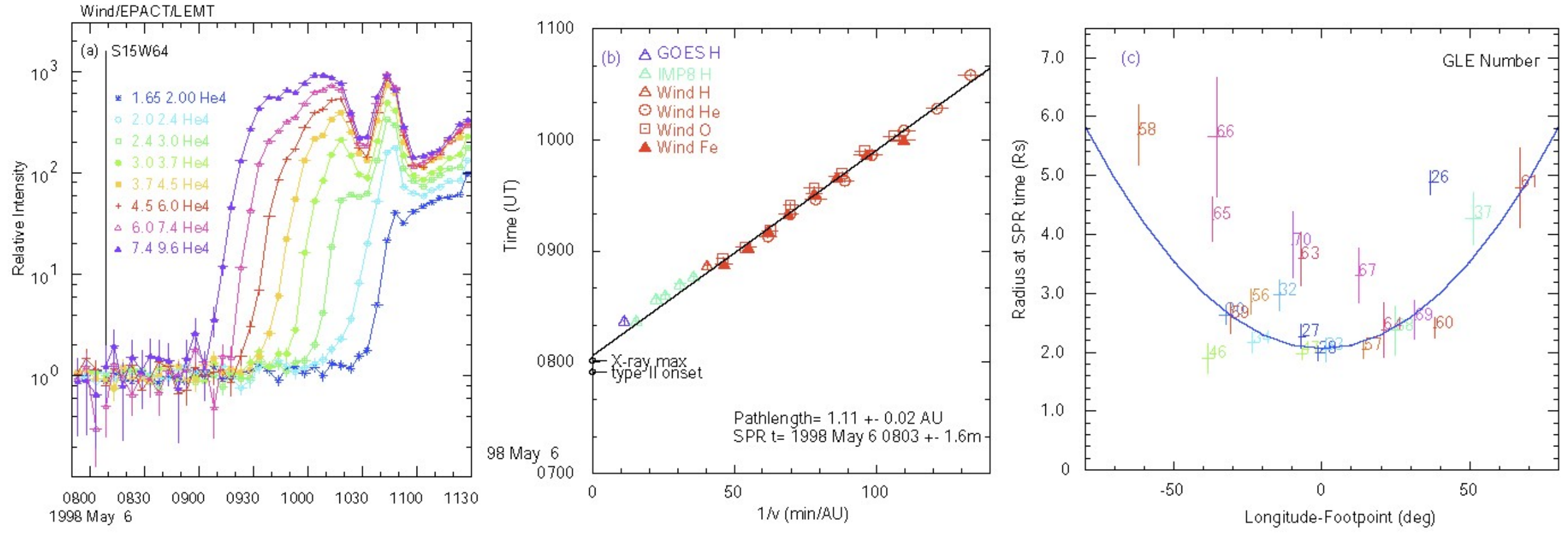

**Fig. 6**(**a**) shows onset times of He intensities at the listed energies (in MeV amu$^{-1}$) in a GLE event while (**b**) shows a plot of corresponding onsets vs. $v^{-1}$ where the slope is the pathlength and the onset is the SPR time at the Sun. (**c**) shows radius vs. connection longitude for several GLEs.

Similar "velocity dispersion" analysis is commonly employed in SEP event onsets [e.g. Kouloumvakos 2025], but is sometimes subverted, when spacecraft intercept magnetic flux tubes with events already in progress, in a complex magnetic environment, or in multiple events. At energies below 1 MeV amu$^{-1}$ it takes over 2 hrs. for the ions to reach 1 AU – enough time for the magnetic connection to change [Ding et al. 2025]. In small events, however, low energies can show velocity dispersion while high energies show the inverse, interpreted either as the slow evolution of shock acceleration [Li et al. 2025a], or perhaps just as a slowly improving magnetic connection from the flank to the nose of the shock [Chan et al 2025; Li et al. 2025b; Kouloumvakos et al. 2026]. In contrast, $^{3}$He-rich events show velocity dispersion [Reames, von Rosenvinge and Lin 1985; Mason et al. 2025] suggesting fast acceleration in magnetic reconnection events in jets and flares.

Between the onset and the ESP event, *large* gradual SEP events form a spectral plateau region with low-energy spectra flattened at the "streaming limit" [Reames and Ng 1998; Ng, Reames, and Tylka 2012], constant within a factor of ~2, shown for several events in Fig. 7a [Reames 1990]. At the streaming limit, an increase in SEPs at the source would soon grow more waves near the shock and choke off any additional flow [Ng, Reames, and Tylka 2003, 2012; Ng 2014; Reames and Ng 2014]. Figure 7b shows how ion spectra of H and O are flattened at the streaming limit at low energy [Reames and Ng 2010] and Fig. 7c shows the effect of intense energetic protons in events that are and are not flattened [Ng, Reames, and Tylka 2012]. Self-amplified waves control the early spectra of large gradual SEP events. The generation of upstream waves that produce the streaming limit (Fig. 7) decouples the upstream and downstream SEP intensities, i.e. it is not possible to predict the ESP or reservoir intensities from upstream intensities.

Sometimes there are efforts to fit the average SEP spectra to a kappa distribution, but there is no physical basis for this. In intense events the shock spectrum is basically a power law and the low-energy rollover is caused by streaming-limited particle transport through proton-amplified waves [Reames and Ng 2010; Ng, Reames, and Tylka 2003, 2012]. Lee et al. [2024] did appropriately use the kappa distribution for the ambient solar wind.

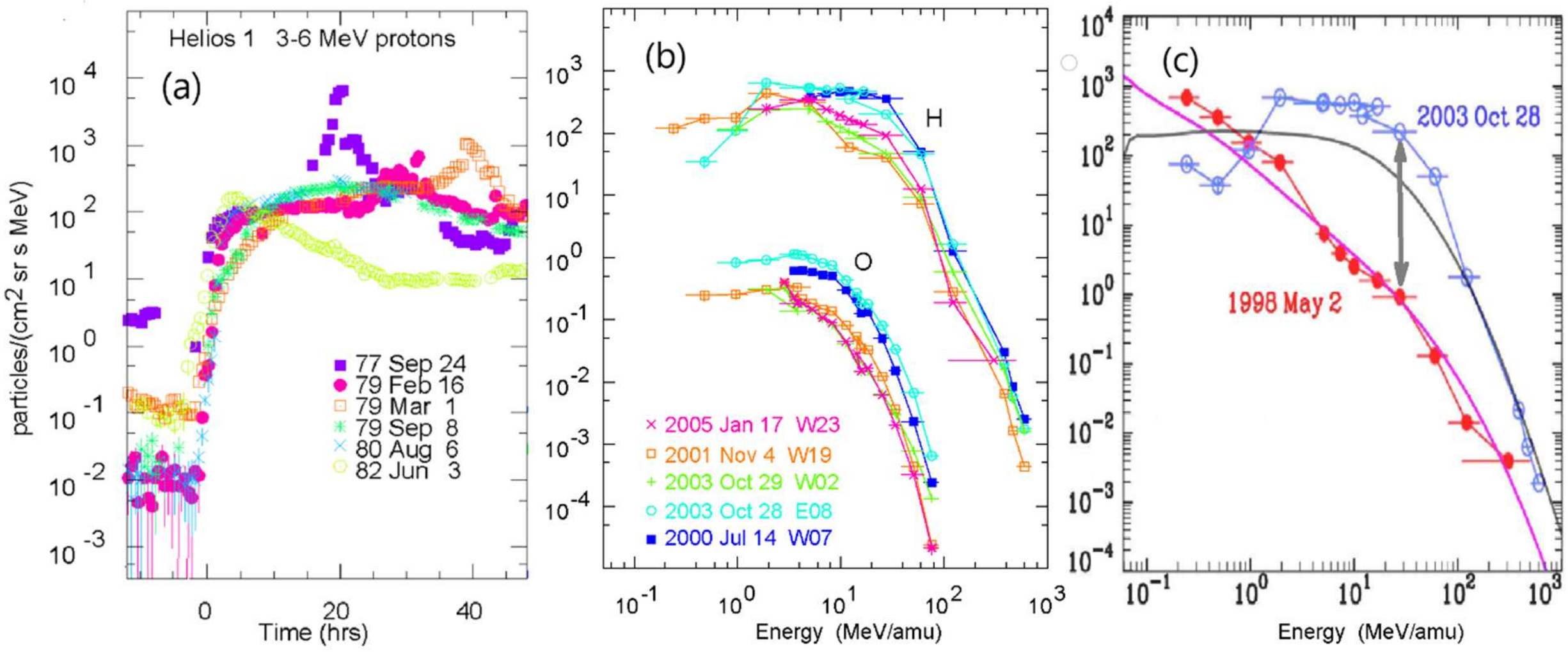

**Fig. 7** (**a**) Superposed time profiles of 3 – 6 MeV protons for the listed SEP events show early plateau intensities at the "streaming limit" at about 200 units, rising much higher at the shock. (**b**) Shows plateau proton and O spectra of the listed events, flattened by self-generated waves at low energies. (**c**) Compares proton spectra in events with greatly different intermediate-energy intensities (gray arrow) needed for growing waves with the rose and gray spectral fits from theory of Ng, Reames, and Tylka [2003, 2012] to the red and blue event data, respectively.

## 5.3 Downstream: Reservoirs and Long-Duration Gamma Rays

The region sunward of the shocks and ESP events in large gradual SEP events has been described as the "reservoir" [Roelof et al. 1992] that fills with spectra that are uniform in the space bounded by the ESP event or waves generated by outflowing particles. Roelof et al. [1992] found nearly equal fluxes of ions and electrons after a large SEP event, between *Ulysses* at 2.5 AU and IMP-8 near Earth. McKibben [1972] had found equal intensities of ~20 MeV protons spanning ~180° in longitude usi2ng IMP-8 with *Pioneer* 6 and 7 and Lario [2010] found reservoirs of electrons between *Ulysses* and ACE extending to high latitudes.

As this contained volume expands, the intensities decrease adiabatically, preserving the shape of the particle spectra. An example of spectral invariance [Reames, Kahler, and Ng 1997; Reames 2013] is shown in Fig. 8 where the spacecraft locations in Fig. 8a are drawn as moving inward (while the CME actually expands outward) tracing the proton intensities vs. time in Fig. 8b with shock crossing times shown. Full spectra at a single time *R* at the three spacecraft positions is shown in Fig. 8c. Reservoirs have spatially uniform intensity spectra and abundances throughout, and all intensities decrease adiabatically as the volume of containment expands.

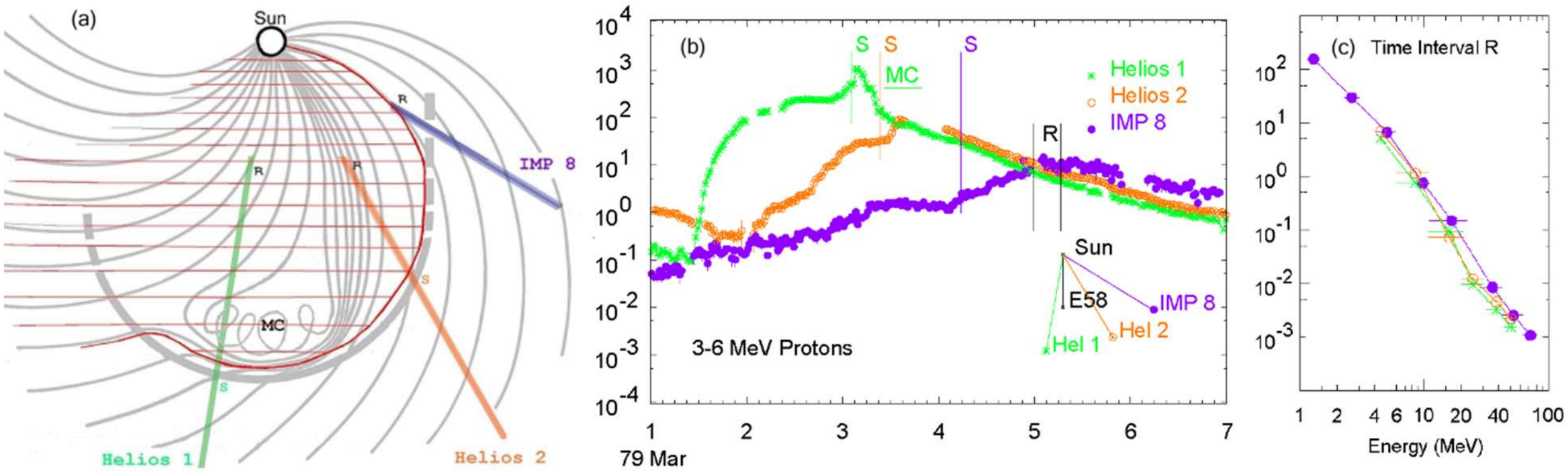

**Fig. 8** (**a**) Cartoon shows three spacecraft moving inward, as the shock (S) and the driving magnetic cloud (MC) actually expand outward, with (**b**) the intensities of 3 – 6 MeV protons vs. time shown at each spacecraft and (**c**) the full spectra at each spacecraft are the same at three different reservoir positions at time R.

The reservoir may rapidly trap high-energy (>300 MeV) protons early, behind the shock and the upstream waves. These protons, and especially any trapped in large magnetic loops, precipitate into the footpoints, producing $\pi^0$ mesons that decay into two (~100 MeV) $\gamma$-rays. Krittinatham and Roffolo [2009] calculate that 1 GeV protons can easily be trapped in a tokamak-like configuration in a flux rope for several hours. These are the long-duration $\gamma$-rays, which continue long after the X-rays from the associated flare and have been seen for many years [Ryan 2000; Chupp and Ryan 2009; Share et al. 2018; Kouloumvakos et. al. 2020; Share 2026]. Recent studies of behind-the-limb flares [Warmuth et al. 2025] show that the duration of the γ-ray events matches that of the shock-driven type-II radio bursts, consistent with the general behavior [Gopalswamy et al. 2025a, b] and found theoretically [Vainio et al. 2026].

Two long-duration $\gamma$-ray events, shown in Fig. 9 [Share and Murphy 2026], are excellent examples of the short, but clear, onset delays of the $\pi^0$-decay radiation.

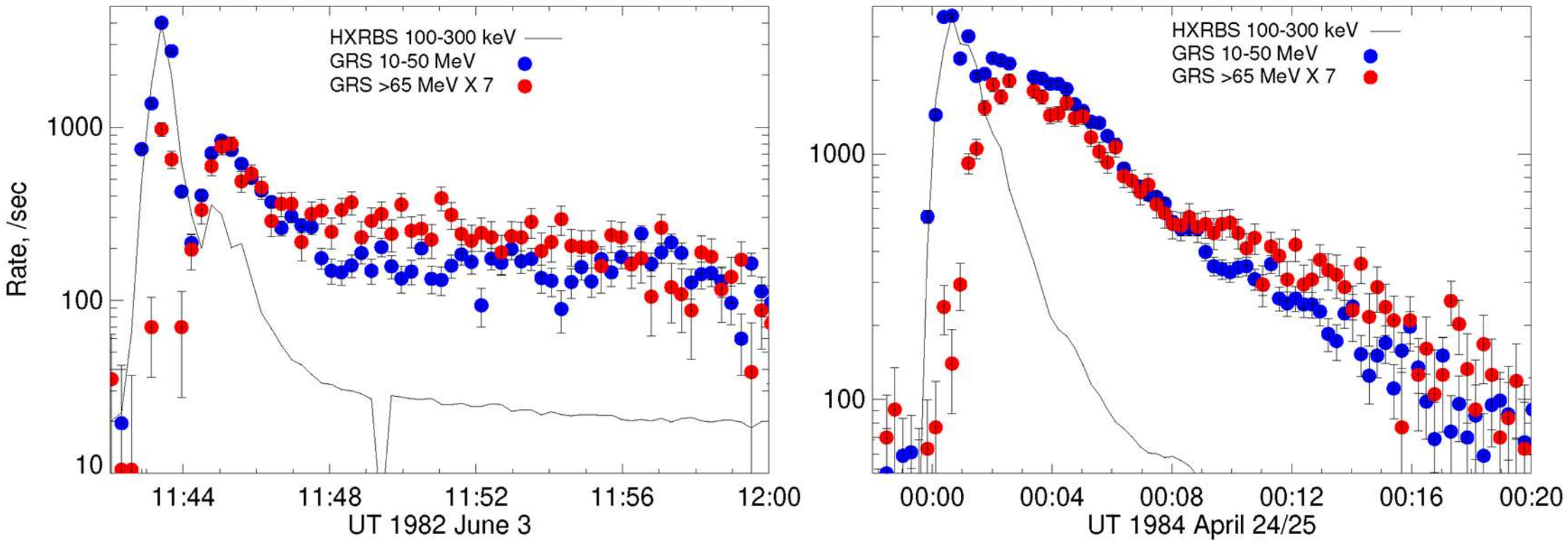

**Fig. 9** shows rates vs. time of the listed X-ray and $\gamma$-ray emissions in two long-duration $\gamma$-ray events. The 100 – 300 keV rates show the impulsive phase. Seven times the rate of >65 MeV $\gamma$-rays will equal the 10-50 MeV rate for the $\gamma$-rays from $\pi^0$ decay. Separate late phases suggest shock-accelerated high-energy protons clearly separated from the impulsive flare. The shock-accelerated protons are still there even in events with no delay.

The generation of upstream waves that produces the streaming limit (Fig. 7) decouples the SEP intensities upstream and downstream so the delayed $\gamma$-rays become unrelated to the bounded SEPs escaping upstream.

## 5.4 Multiple CMEs: More Enhancements from Seeding

Gopalswamy et al. [2002] suggested that interacting CMEs, where a faster CME overtakes a slower one within ~20 solar radii, might produce higher SEP intensities than a single fast CME would accelerate alone. Richardson et al. [2003] objected that these results must be spurious since most of the energetic SEPs are accelerated early by the fast CME, prior to any interaction. Gopalswamy et al. [2004] divided 57 well-observed large SEP events into three categories: 23 events with a preceding CME (P), 20 events with no preceding CME (NP), and other (O) events that include CMEs traversing high-density streamers. The three categories show different correlations with CME speed, the P and O events showed the highest proton intensities at high CME speeds, suggesting the highest populations of seed particles. We note that there is likely to be a reservoir [Sect. 5.3; Reames 2013] of relatively intense SEPs behind the first CME, *all the way to the Sun*, to provide an additional population of energetic seed particles at the fast shock in the second SEP events.

Figure 10 shows a calculation of the time evolution of the radial distribution of 2.6 MeV protons from Ng, Reames, and Tylka [2003]. The upper bound of the curves maps out the radial distribution of the streaming limit, discussed in Section 5.2 while the flat extension of each curve to lower radius approximates the reservoir level at each time. While the ESP peak is not correctly modeled, the scaling is still appropriate. Note that the reservoir level is quite high initially, i.e inside 20 $R_S$ (0.1 AU); it is these particles that serve as seed particles when a shock from a second CME samples them. Unfortunately, seed particles from the earlier CME are not as distinguishable as those from earlier impulsive SEPs, but they can boost the resulting SEP intensities, nevertheless.

**Fig. 10** Shows the calculated radial distribution of the calculated intensities of 2.6 MeV protons at various listed times [Ng, Reames, and Tylka 2003]. The radial dependence of the streaming limit is indicated as is the time variation of the reservoir level which would serve as an additional seed population for a second shock, especially within the first 20 $R_s$ (5 hr).

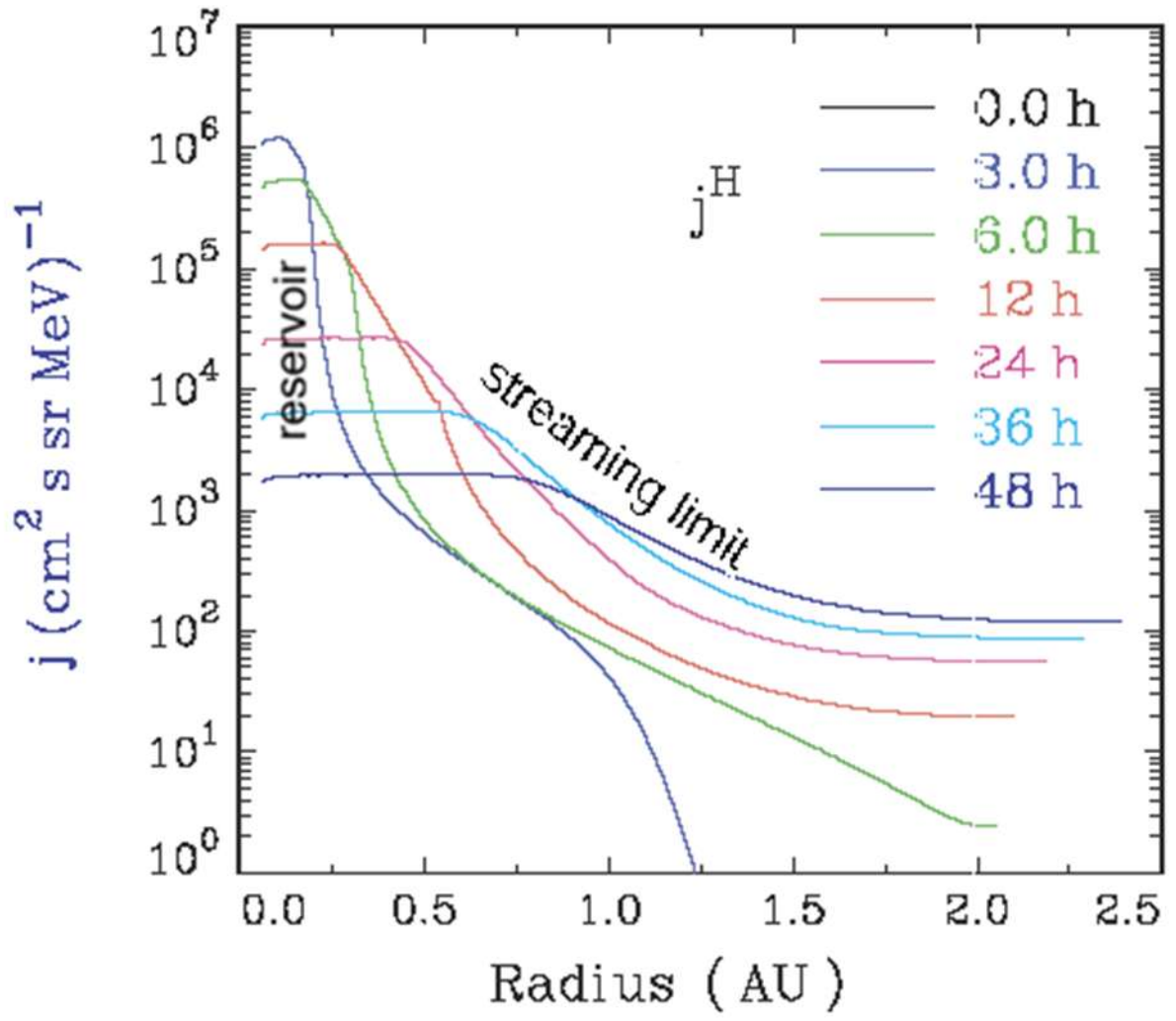

## 5.5 Bidirectional Flows

Long ago, Piddington [1958] and especially Gold [1958] suggested that magnetic loops drawn out from the Sun would remain attached to form a topologically closed bottle (often called the "Gold bottle"), since Maxwell's equations require that all magnetic field lines are closed, during ejections we now know of as CMEs. Rao, McCracken and Bukata [1967] first measured four periods of bidirectional anisotropies of 10 MeV protons which they argued were not consistent with the Gold bottle. Subsequent authors [Marsden et al. 1987; Gosling et al. 1987] also considered the "plasmoid" alternative where magnetic reconnection could result in ejection of a closed-field loop. Kahler and Reames [1991] noted that bidirectional flows are not always associated with CMEs and Reames [2002] noted that the detached loop in two-dimensional cross sections was actually a spiral flux rope connected to the Sun in three dimensions.

Figure 11 from Tan et al. [2013] shows electrons arriving along the field from the Sun and an hour later coming from the opposite direction, possibly reflected ~1.5 AU further on (40 keV electrons travel 1 AU in ~20 min), to become a bidirectional electron flow.

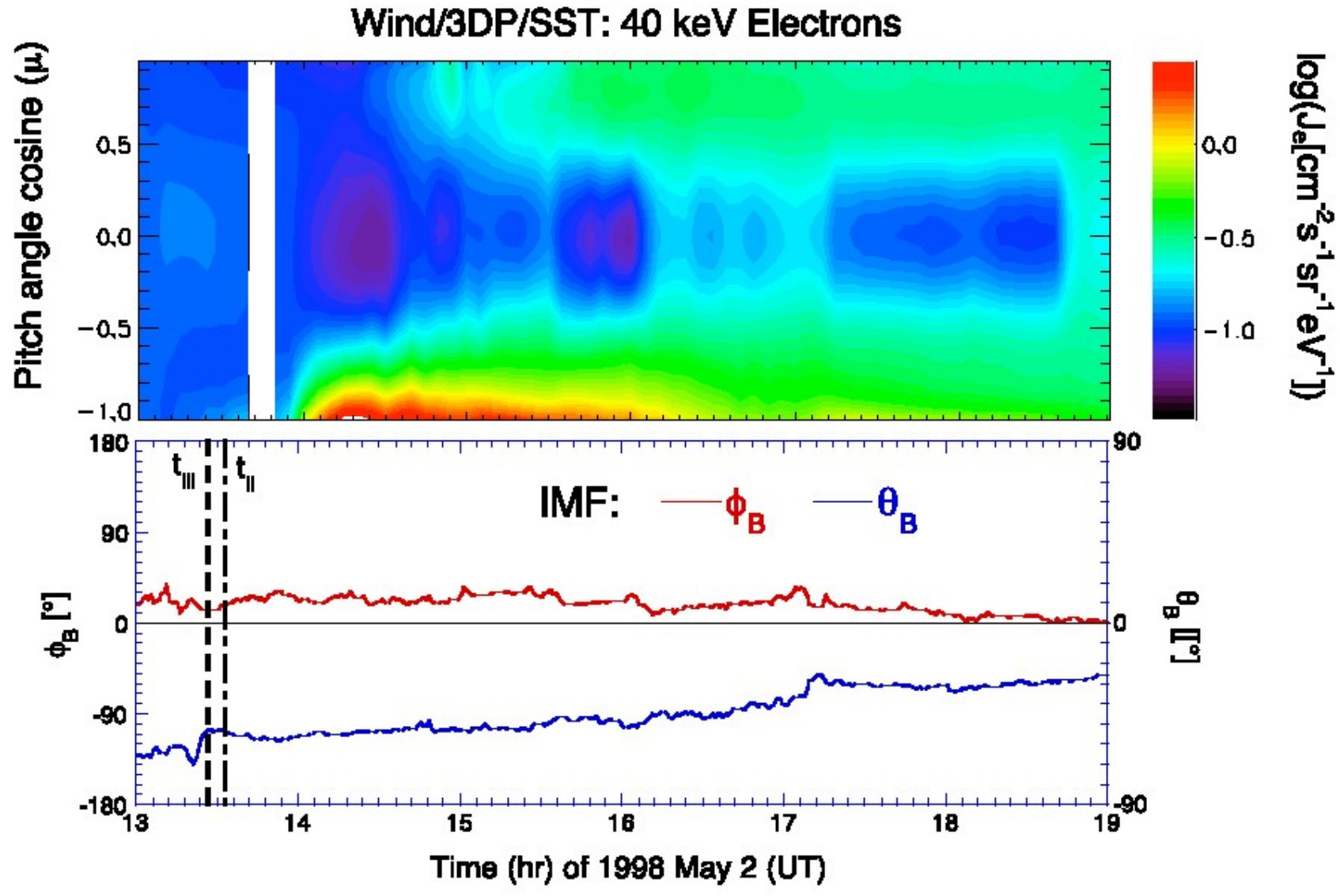

**Fig. 11** The time variation of the pitch-angle distribution of 40 keV electrons in the 2 May 1998 event is shown in the upper panel with the magnetic field direction below. The electrons first arrive from the Sun at μ ≈ -1 at 14:00 UT, then reflected electrons arrive at μ ≈ +1 at ≈15:00 UT. Times of type III and type II radio emission are noted in the lower panel.

A list of over 1000 bidirectional ion flow periods >4 hrs of ~1 MeV protons near Earth between 1973 and 1991 is has been compiled by Richardson and Reames [1993], while Richardson [1994] provided a similar list for *Helios 1* and *2* from 0.3 – 1.0 AU between 1974 and 1984.

Recently Ding et al. [2025] have analyzed events in terms of propagation of SEPs on loops inside a flux rope, also considering shocks that can intercept both ends of a loop.

# 6 Reference Abundances the Corona and FIP

Large SEPs have the advantage of measuring ~20 elements at a time so it is possible to measure full FIP patterns, not just high and low FIP, and to decouple other effects. When it became possible to compare the abundances of elements in SEPs, relative to reference abundances, such as those in the solar photosphere, two dependences were found [Meyer 1985; Breneman and Stone, 1985]: (1) an average difference was dependent upon the first ionization potential (FIP) of each element and (2) an event-to-event variation was found to have a power-law dependence on the atomic mass-to-charge ratio *A/Q* of each ion.

This FIP-dependence had been anticipated long in advance [e.g. Webber 1974] that could separate low-FIP (≤10 eV) ions from high-FIP neutral atoms as they diffuse across the chromosphere from the photosphere to the corona. Present theory [Laming 2015] suggests that elements that are ionized, such as Mg, Si, and Fe are boosted by the ponderomotive force of Alfvén waves, while initially-neutral atoms such as O, Ne, and He are not, making a relative enhancement ≈ 3× at low FIP. All elements are then easily ionized as they enter the hot ~1 MK corona.

The power-law dependence upon *A/Q* occurs since abundances of ions that are compared at a given velocity (or MeV $amu^{-1}$) scatter as their magnetic rigidity which varies as *A/Q* times velocity. When high-rigidity ions scatter less, they can escape the area of observation more easily, suppressing the abundance Fe/O, for example, or, with greater scattering, they can escape more easily from the source to the observation, initially enhancing the observed Fe/O. Fe/O has been a proxy for the power law in *A/Q*. The power-law dependence of *A/Q* became more evident [Breneman and Stone 1985; Tylka et al. 2000] when measurements of ionization states *Q* became possible [Luhn et al. 1984, 1987]. For impulsive SEP events, a systematic enhancement with atomic number *Z* was known [Mason et al. 1986] but the knowledge of a *A/Q*-dependence began to emerge when it became clearer that the measured charges had been stripped away after acceleration [Reames, Meyer, and von Rosenvinge 1994; DiFabio et al. 2008]. Since the SEP1 ions are stripped *after* acceleration (e.g. to $Q_{Fe} \approx 20$), all of these shock-reaccelerated ions in SEP2 and SEP3 events also have $Q_{Fe} \approx 20$, even in SEP3 GLEs [Mewaldt et al. 2012; Reames 2024b]. However, power-law fits to these high-*Z* abundances respond to the temperature of the original abundance-determining source.

The underlying FIP dependence of SEP abundances seem to be the same in both impulsive and gradual SEP events. However, this FIP pattern of SEPs was found [Reames, Richardson, and Barbier 1991; Mewaldt et al. 2002; Reames 2018a] to differs from that of the solar wind and of energetic particles accelerated from the solar wind at corotating-interaction-region (CIR) shocks where high-speed solar wind streams intercept slow wind emitted earlier in the solar rotation. CIR shocks, which often form beyond 1 AU, likely accelerate ions from the local solar wind. In contrast, a study of SEP abundances in fast and slow wind showed no differences [Kahler, Tylka, and Reames 2009], i.e. SEPs do not depend upon the solar wind.

Figure 12a shows the FIP patterns for SEPs compared with the theory for closed loops [Laming 2015; Laming et al. 2019] where the resonance of Alfvén waves with the loop length tends to suppress elements C, P, and S, in the blue band more than the corresponding element in the solar wind, presumably derived from more-open field lines and shown in Fig. 12b.

**Fig. 12** (**a**) Shows SEP element abundances (solid blue circles) relative to those of the photosphere vs. the FIP of each element, along with the theoretical values (open red circles) for closed magnetic loops. (**b**) Shows the corresponding values for the solar wind (solid blue circles), for CIRs (solid blue squares), and average theoretical values for the solar wind (open red circles).

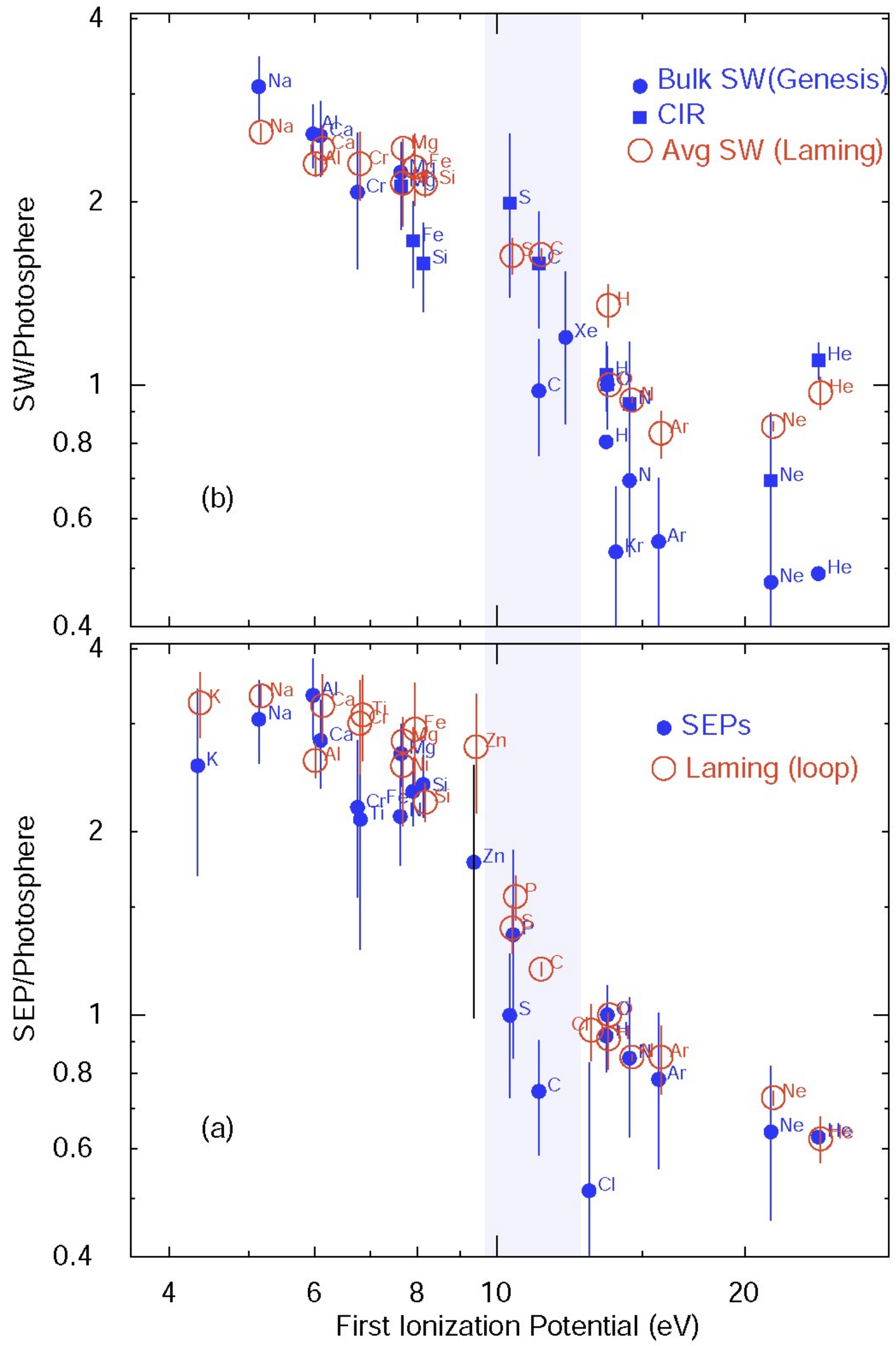

In Fig. 12 and Table 1, the element abundances in the photosphere are derived from Bergemann, Lodders, and Palma [2025], those of the bulk solar wind from the Genesis sample experiment [Heber et al. 2021], while the SEP and CIR abundances are the same as those previously tabulated by Reames [2020a, 2021a, b], which used earlier photospheric and solar-wind abundances for comparison.

**Table 1** Photospheric, Reference SEPs, CIR, and Solar Wind Abundances.

| | Z | FIP [eV] | Photosphere[1] | SEPs[2] | CIRs[3] | Bulk SW Genesis[4] |
|---|---|---|---|---|---|---|
| H | 1 | 13.6 | (1.74±0.02) ×10^6 | (≈1.6±0.2) ×10^6 | (1.81±0.24) ×10^6 | (1.40±0.012) ×10^6 |
| He | 2 | 24.6 | 145000±4000 | 91000±5000 | 159000±10000 | 70900±190 |
| C | 6 | 11.3 | 562±120 | 420±10 | 890±36 | 548±27 |
| N | 7 | 14.5 | 151±38 | 128±8 | 140±14 | 105±2.6 |
| O | 8 | 13.6 | 1000±120 | 1000±10 | 1000±37 | 1000±100 |
| Ne | 10 | 21.6 | 245±68 | 157±10 | 170±16 | 116±.51 |
| Na | 11 | 5.1 | 3.39±0.39 | 10.4±1.1 | – | 10.5±.34 |
| Mg | 12 | 7.6 | 66.1±7.6 | 178±4 | 140±14 | 148±3.4 |
| Al | 13 | 6.0 | 4.68±0.54 | 15.7±1.6 | – | 12.1±0.43 |
| Si | 14 | 8.2 | 63.1±7.3 | 151±4 | 100±12 | – |
| P | 15 | 10.5 | 0.48±0.13 | 0.65±0.17 | – | – |
| S | 16 | 10.4 | 25.1±6.4 | 25±2 | 50±8 | – |
| Cl | 17 | 13.0 | 0.47±0.22 | 0.24±0.1 | – | – |
| Ar | 18 | 15.8 | 5.5±1.5 | 4.3±0.4 | – | 3.03±0.015 |
| K | 19 | 4.3 | 0.214±0.044 | 0.55±0.15 | – | – |
| Ca | 20 | 6.1 | 3.89±0.54 | 11±1 | – | 10.0±0.34 |
| Ti | 22 | 6.8 | 0.162±0.041 | 0.34±0.1 | – | – |
| Cr | 24 | 6.8 | 0.96±0.24 | 2.1±0.3 | – | 1.98±0.043 |
| Fe | 26 | 7.9 | 56.2±6.5 | 131±6 | 97±11 | – |
| Ni | 28 | 7.6 | 3.02±0.42 | 6.4±0.6 | – | – |
| Zn | 30 | 9.4 | 0.062±0.016 | 0.11±0.04 | – | – |
| Kr | 36 | 13.99 | 0.00355±0.00098 | – | – | 0.00188±0.00010 |
| Xe | 54 | 12.13 | (3.47±0.96) ×10^-4 | – | – | (4.16±0.25) ×10^-4 |
| Se–Zr | 34–40 | – | – | 0.04±0.01 | – | – |
| Sn–Ba | 50–56 | – | – | 0.0066±0.001 | – | – |
| Os–Pb | 76–82 | – | – | 0.0007±0.0003 | – | – |

[1] Bergemann, Lodders, and Palma [2025].
[2] Reames [1995a, 2014, 2020, 2021a].
[3] Reames, Richardson, and Barbier [1991]; Reames [1995a, 2020,2021a].
[4] Heber et al. [2021].

Earlier solar-wind abundances reported by Bochsler [2009] were in better agreement with the CIR abundances, with enhanced abundances of both He and C, relative to those in SEPs, being interpreted [see Reames 2018a, 2020] as differences in open- vs. closed-field FIP fractionation for the solar-wind and SEP sources, respectively. The CIR abundances for He and C also fit the Laming [2015] theory for open field lines. However, the recent measurements of abundances of solar-wind ions captured in the Si targets of the Genesis experiment [Heber et al. 2021] are much closer to the SEP values, with C now lying close to the photospheric value, as can be seen in Table 1. Thus, the Genesis results seem to erase the longstanding FIP-dependent abundance differences between SEPs and the solar wind [Reames, Richardson, and Barbier 1991; Mewaldt et al.2002; Schmelz et al. 2012; Reames 2018a]. It is probably important to distinguish fast and slow wind, especially for the C abundance, rather than using bulk flow.

When the SEP FIP abundances are compared with maps of FIP bias in the corona [Brooks et al. 2016; Reames 2021] they generally show active regions as SEP sources.

There is recent evidence [Mzerguat et al. 2026; Orlovskij, To, and Long, 2026] that abundances of elements on the boundary between low and high FIP, like S, are dependent upon the magnetic fields in plumes. Thus, FIP-bias as measured by Si/S generally does not reflect that measured by Si/O. The suppressed abundances of S and C in SEPs in Figure 12a might have a similar origin. Perhaps the high fields in sunspots, which generate high coronal loops, populate them with abundances with suppressed S and C and make them available for acceleration by coronal shocks.

# 7 Summary

There are two primary sites of SEP acceleration: (A) at open/closed-field magnetic reconnection sites in solar jets/flares and (B) at CME-driven shock waves

- ✓ (A1) Reconnection produces SEP1 element abundance enhancements in solar jets that are approximately increasing power-laws in *A/Q*, where *Q* depends upon the source electron temperature, usually ~2.5 MK.
- ✓ (A2) Resonant (EMIC) wave absorption in jets greatly enhances $^3$He/$^4$He: waves are damped at $^1$H and $^4$He gyrofrequencies but rare $^3$He continues to absorb them. Heavier ions may be enhanced through second-harmonic absorption, especially at suprathermal energies in small events, e.g. at cool (~1.5 MK) jets near coronal holes.
- ✓ (A3) Solar jets also produce copious electrons that magnetically focus into the beams that generate type-III radio bursts as they stream out from the Sun.
- ✓ (A4) Blowout jets may eject CMEs fast enough (>500 km s$^{-1}$) to drive shocks that reaccelerate the SEP1 seed ions, producing SEP2 events.
- ✓ (A5) Reconnections in only-closed loops produce flares, where the particle energy is magnetically trapped and is dissipated as heat and light (e.g. X-rays).
- ✓ (B1) Wide, fast (~1000 km s$^{-1}$), CME-driven shock waves accelerate ions from the ambient coronal plasma (SEP4 events). Wave generation forms an ESP trapping region (spike) around the shock which often can survive to 1 AU and beyond.
- ✓ (B2) Some active regions provide a continuous stream of multiple impulsive SEP1 jets that provide additional pre-accelerated seed particles for acceleration by a wide fast shock. These are defined as a SEP3 event.
- ✓ (B3) Earliest arriving ions usually show velocity dispersion and higher-rigidity ions may leak away and deplete first in modest events.
- ✓ (B4) In intense events early wave generation can limit outflow to the streaming limit and flatten low-energy spectra of ions. Higher-rigidity ions (e.g. Fe) become enhanced early, since they traverse the waves more easily. Thus, they become depleted at and behind the shock.

- ✓ (B5) In large SEP events, the region between the shock and the Sun, left behind and fed by the expanding ESP, becomes a nearly-uniform reservoir of magnetically-trapped, invariant spectra that decrease adiabatically as the volume of the region expands. Reservoirs provide sustained precipitation for long-duration solar γ-ray events and seeds for multi-shock events.

The *A/Q*-dependence of abundances at constant velocity is a useful tool to distinguish rigidity and velocity dependences. Power-law fits vs. *A/Q* allow estimates of the charge states and temperature of the original coronal ion source. The reference abundances allow study of the FIP fractionation in the coronal source region.

**Acknowledgements** The author thanks Gerry Share and Ed Cliver for helpful discussions and for comments on this manuscript.

**Competing Interests** The author declares he has no conflicts of interest.